# Sparse Median Graphs Estimation in a High Dimensional Semiparametric Model

Fang Han[*]   and   Han Liu[†]   and   Brian Caffo [‡]

**Abstract**

In this manuscript a unified framework for conducting inference on complex aggregated data in high dimensional settings is proposed. The data are assumed to be a collection of multiple non-Gaussian realizations with underlying undirected graphical structures. Utilizing the concept of median graphs in summarizing the commonality across these graphical structures, a novel semiparametric approach to modeling such complex aggregated data is provided along with robust estimation of the median graph, which is assumed to be sparse. The estimator is proved to be consistent in graph recovery and an upper bound on the rate of convergence is given. Experiments on both synthetic and real datasets are conducted to illustrate the empirical usefulness of the proposed models and methods.

## 1   Introduction

Undirected graphs provide a powerful tool for understanding the interrelationships among random variables. Given a random vector, $\boldsymbol{X} = (X_1, \ldots, X_d)^T \in \mathbb{R}^d$, the associated conditional independence graph, say $\mathcal{G} \in \{0,1\}^{d \times d}$, is the undirected binary graph so that the entry $\mathcal{G}_{jk}$ (for $j \neq k$) is equal to 0 if and only if $X_j$ is conditionally independent of $X_k$ given the remaining variables, $\{X_{\setminus\{j,k\}}\}$. For estimation, it is typically assumed that there are $n$ independent and identically distributed realizations of $\boldsymbol{X}$ to infer independence relationships, and thus the associated graph, $\mathcal{G}$.

---

[*]Department of Biostatistics, Johns Hopkins University, Baltimore, MD 21205, USA; e-mail: fhan@jhsph.edu

[†]Department of Operations Research and Financial Engineering, Princeton University, Princeton, NJ 08544, USA; e-mail: hanliu@princeton.edu

[‡]Department of Biostatistics, Johns Hopkins University, Baltimore, MD 21205, USA; e-mail: bcaffo@jhsph.edu

When $\boldsymbol{X} \sim N_d(\boldsymbol{\mu}, \boldsymbol{\Sigma})$ has a multivariate distribution with mean $\boldsymbol{\mu}$ and covariance matrix $\boldsymbol{\Sigma}$, one obtains the key observation that the non-zero entries of the so-called concentration matrix $\boldsymbol{\Omega} := \boldsymbol{\Sigma}^{-1}$, otherwise known as its sparsity pattern, encodes the conditional independence structure of $\boldsymbol{X}$ and hence defines the graph $\mathcal{G}$ (Dempster, 1972). In other words, $\mathcal{G}_{jk} = I(\boldsymbol{\Omega}_{jk} \neq 0)$, where $I(\cdot)$ is an indicator function and $\mathcal{G}_{jk}$ indicates whether an edge connects nodes $j$ and $k$ in the graph. Estimation of the concentration matrix becomes problematic in high dimensional settings where $d > n$, thus leading to an active collection of research utilizing sparsity constraints to obtain identifiability (see Friedman et al., 2007; Banerjee et al., 2008; Li and Toh, 2010; Scheinberg et al., 2010; Hsieh et al., 2011; Rothman et al., 2008; Ravikumar et al., 2009; Lam and Fan, 2009; Peng et al., 2009; Meinshausen and Bühlmann, 2006; Yuan, 2010; Cai et al., 2011; Liu and Luo, 2012, for example).

However, these manuscripts all assumed that the object of inference is a single graph estimated from a single set of realizations of $\boldsymbol{X}$. In contrast, little work exists on estimation and inference from a population of graphs. Such a setting arises frequently in the sometimes controversial and rapidly evolving arenas of image- and electrophysiologically- based estimates of functional and structural brain connectivity (Friston, 2011; Horwitz et al., 2003; Fingelkurts et al., 2005; Rubinov and Sporns, 2010; Bullmore and Sporns, 2009). Here each subject-specific graph is an estimate of subject-specific brain connectivity. To date, no theoretically justified definition for population graphs exists.

In addition, frequently the assumption that the data are independently and identically drawn from a Gaussian distribution is too strong. Recently, Gaussian assumptions were relaxed via the so-called *nonparanormal* distribution family (Liu et al., 2009). A random vector, $\boldsymbol{X}$, is said to be nonparanormally distributed if, after an unspecified monotone transformation, it is Gaussian distributed. Moreover, an optimal rate in graph recovery is obtained utilizing the rank-based estimator Kendall's tau (Liu et al., 2012). On the other hand, however, little has been done in high dimensional graph estimation when the data are actually not identically drawn from a certain distribution.

This manuscript investigates a specific non-iid setting where the data arise separate multiple datasets, each assumed to be distributed according to a different distribution. This idea is central in fields, such as epidemiology, where population summaries, are desired over a collections of independently but not identically distributed data sets. A canonical example is the common odds ratio estimate from a collection individual odds ratios (see Liu and Agresti, 1996, for example). In the motivating application, each dataset is a seed-based or region of interest summary of functional magnetic resonance imaging (fMRI) scans where a graphical representation of brain connectivity is of interest. The proposed approach does not assume a common underlying graph for each subject. Instead, the population graph defined is a summary, looking at commonalities in graphical structure across a population of heterogeneous graphs. Thus it is proposed that under the presumption of variation in brain graphical network structure, the investigation of a population graph is of conceptual and practical interest, especially when comparing population graphs across clinical diagnoses.

To best summarize the information from aggregated network datasets, the idea of "median graphs" from the pattern recognition field (Bunke and Shearer, 1998; Jiang et al., 2001) is employed. However, it is herein extended to *sparse median graphs*. A sparse median graph is defined as the sparse graph that has the smallest sum of Hamming distances to all graphs in a given sample. Combined with the strength of the nonparanormal modeling, a new method for estimating sparse median graphs is proposed. It is then proven that the obtained estimator is consistent and the upper bound on the convergence rate with respect to the Hamming distance is established.

The rest of the paper is organized as follows. In Section 2 notation is introduced and the nonparanormal distribution and rank-based estimators are reviewed. In Section 3, the model is introduced and the definition of sparse median graphs is given. In Section 4, rank-based estimation procedures are provided. In Section 5, the theoretical properties of the proposed procedure for graph recovery are given. Section 6 demonstrates experimental results on both synthetic and real-world datasets to back up our theoretical results. Discussions are provided in the last section.

## 2 Background

Let $\mathbf{M} = [M_{jk}] \in \mathbb{R}^{d \times d}$ and $\boldsymbol{v} = (v_1, ..., v_d)^T \in \mathbb{R}^d$. Let the subvector of $\boldsymbol{v}$ with entries indexed by set $I$ be denoted by $\boldsymbol{v}_I$. Similarly, let the submatrix of $\mathbf{M}$ with rows indexed by set $I$ and columns indexed by set $J$ be denoted by $\mathbf{M}_{IJ}$. Let $\mathbf{M}_{I*}$ and $\mathbf{M}_{*J}$ be the submatrix of $\mathbf{M}$ with rows in $I$, and the submatrix of $\mathbf{M}$ with columns in $J$. For $0 < q < \infty$, define the $\ell_0, \ell_q$ and $\ell_\infty$ vector norms as follows:

$$||\boldsymbol{v}||_0 := \text{card}\{\text{supp}(\boldsymbol{v})\}, \quad ||\boldsymbol{v}||_q := \left(\sum_{i=1}^d |v_i|^q\right)^{1/q} \quad \text{and} \quad ||\boldsymbol{v}||_\infty := \max_{1 \le i \le d} |v_i|.$$

Likewise, for matrix norms:

$$||\mathbf{M}||_q := \max_{||\boldsymbol{v}||_q = 1} ||\mathbf{M}\boldsymbol{v}||_q, \quad ||\mathbf{M}||_{\max} := \max\{|M_{ij}|\}, \quad \text{and } ||\mathbf{M}||_H := \sum_{j > k} I(\mathbf{M}_{jk} \ne 0).$$

We define diag(**M**) to be an diagonal matrix with diagonal values same as that of **M** and off-diagonal values to be zero.

## 2.1 The Nonparanormal

Liu et al. (2009) and Liu et al. (2012) showed that the Gaussian graphical model can be relaxed to the nonparanormal graphical model without significant loss of inference power when the data are Gaussian distributed and with significant gain of inference power when it is not. This observation plays an role in our proposed model for relaxing the Gaussian assumption. In this section, the nonparanormal distribution family is introduced with the corresponding graphical model, following definitions in Liu et al. (2012).

**Definition 2.1** (The nonparanormal). *Let $f = \{f_j\}_{j=1}^d$ be a set of univariate strictly increasing functions. A d-dimensional random vector $\boldsymbol{X} = (X_1, \ldots, X_d)^T$ is said to follow a nonparanormal distribution, denoted $NPN_d(\boldsymbol{\Sigma}, f)$, if and only if*

$$f(\boldsymbol{X}) := \{f_1(X_1), \ldots, f_d(X_d)\}^T \sim N_d(\boldsymbol{0}, \boldsymbol{\Sigma}), \text{ where } \mathrm{diag}(\boldsymbol{\Sigma}) = \mathbf{I}_d,$$

*where $\mathbf{I}_d \in \mathbb{R}^{d \times d}$ is the identity matrix. $\boldsymbol{\Sigma}$ is referred to as the latent correlation matrix and $\boldsymbol{\Omega} := \boldsymbol{\Sigma}^{-1}$ as the latent concentration matrix.*

Although the nonparanormal distribution family is strictly larger than the Gaussian distribution family, Liu et al. (2009) showed that the conditional independence property of the nonparanormal is still encoded in the latent concentration matrix $\boldsymbol{\Omega}$. More specifically, provided that the random vector $\boldsymbol{X} = (X_1, \ldots, X_d)^T$ is nonparanormally distributed (i.e. $\boldsymbol{X} \sim NPN_d(\boldsymbol{\Sigma}, f)$) then

$$X_j \perp X_k \,|\, X_{\setminus j,k} \Leftrightarrow \boldsymbol{\Omega}_{jk} = 0. \tag{2.1}$$

## 2.2 Rank-based Estimator

Liu et al. (2012) and Xue and Zou (2012) exploited the rank-based estimator, Kendall's tau, in inferring the latent concentration matrix $\boldsymbol{\Omega}$ in the nonparanormal family. Let $\boldsymbol{x}_1, \ldots, \boldsymbol{x}_n \in \mathbb{R}^d$, with $\boldsymbol{x}_i = (x_{i1}, \ldots, x_{id})^T$ for $i = 1, \ldots, n$, be $n$ observed data points of a random vector $\boldsymbol{X}$. The Kendall's tau statistic is defined as:

$$\widehat{\tau}_{jk}(\boldsymbol{x}_1, \ldots, \boldsymbol{x}_n) := \frac{2}{n(n-1)} \sum_{1 \leq i < i' \leq n} \mathrm{sign}(x_{ij} - x_{i'j}) \cdot (x_{ik} - x_{i'k}). \tag{2.2}$$

The Kendall's tau statistic is monotone transformation-invariant correlation between the empirical realizations of $X_j$ and $X_k$ for any $j, k \in \{1, \ldots, d\}$. Let $\widehat{\mathbf{R}} = [\widehat{\mathbf{R}}_{jk}] \in \mathbb{R}^{d \times d}$, with

$$\widehat{\mathbf{R}}_{jk} = \sin\left(\frac{\pi}{2}\widehat{\tau}_{jk}(\boldsymbol{x}_1, \ldots, \boldsymbol{x}_n)\right), \tag{2.3}$$

be the Kendall's tau matrix. Liu et al. (2012) showed that if $\boldsymbol{X}$ is nonparanormally distributed, $\widehat{\mathbf{R}}$ is a consistent estimator of the latent correlation matrix $\boldsymbol{\Sigma}$ of $\boldsymbol{X}$ (with respect to element-wise sup norm $||\cdot||_{\max}$), even when the order of $d$ is nearly exponentially larger than $n$.

Since the latent concentration matrix, $\boldsymbol{\Omega} = \boldsymbol{\Sigma}^{-1}$, fully encodes the nonparanormal graphical model and $\widehat{\mathbf{R}}$ is a consistent estimator of $\boldsymbol{\Sigma}$, Kendal's tau is a good estimate in estimating the nonparanormal graphical model, as it directly estimates the latent concentration matrix. Based on the Kendall's tau matrix, Liu et al. (2012) proposed the nonparanormal SKEPTIC by directly plugging $\widehat{\mathbf{R}}$ into any statistical methods in calculating the inverse covariance/correlation matrix. In this paper, we will focus on one particular statistical method, CLIME (Cai et al., 2011). Further details of the non paranormal SKEPTIC are given in Section 4.

# 3 Models and Concepts

## 3.1 Models

In this section the proposed approach for modeling the complex aggregated data is given. Assume that the data are aggregated from multiple datasets, each of which is distributed according to a different nonparanormal distribution.

More specifically, let $\boldsymbol{X}_1, \ldots, \boldsymbol{X}_T$ be $T$ different random vectors with $\boldsymbol{X}_t = (X_{t1}, \ldots, X_{td})^T$ satisfying that

$$\boldsymbol{X}_t \sim NPN_d(\boldsymbol{\Sigma}^t, f^t), \quad \text{for } t = 1, \ldots, T.$$

Let $\boldsymbol{\Theta}^t := [\boldsymbol{\Sigma}^t]^{-1}$. Based on $\boldsymbol{\Theta}^t$, we define $\mathcal{G}^t = [\mathcal{G}_{jk}^t] \in \{0, 1\}^{d \times d}$ where

$$\mathcal{G}_{jk}^t = 0 \text{ if and only if } \boldsymbol{\Theta}_{jk}^t = 0.$$

Via Equation 2.1, $\mathcal{G}^t$ represents the Markov graph associated with $\boldsymbol{X}_t$. In detail, the pair $(j, k)$ such that $\mathcal{G}_{jk}^t \neq 0$ indicates the conditional independence of $X_{tj}$ and $X_{tk}$ given all the rest in $\boldsymbol{X}_t$.

## 3.2 Sparse Median Graphs

In this section, the concept of a *sparse median graph* is introduced, combining the ideas of median graphs from Jiang et al. (2001) and the sparsity concept commonly adopted in high dimensional statistic (Bühlmann and van de Geer, 2011).

Let $d(\cdot) : \{0,1\}^{d \times d} \times \{0,1\}^{d \times d} \to [0, \infty)$ be a distance function on the graph space. Jiang et al. (2001) define the median graph (reproduced in Definition 3.1) as the graph that has the smallest sum of distances to all graphs in a given set.

**Definition 3.1** (Median Graph). *Let $\mathcal{G}^1, \ldots, \mathcal{G}^T$ be $T$ different binary graphs in $\{0,1\}^{d \times d}$, the median graph $\mathcal{G}^*$ is defined by*

$$\mathcal{G}^* := \underset{\mathcal{G} \in \{0,1\}^{d \times d}}{\operatorname{argmin}} \sum_{t=1}^{T} d(\mathcal{G}, \mathcal{G}^t). \tag{3.1}$$

When $T$ is large, $\mathcal{G}^*$ will not be sparse and therefore the resulting median graph may not be interpretable. To attack this problem, consider the concept of a "sparse median graph". The sparse median graph is the graph that has the smallest sum of distances to all graphs in a given set and the non-zero entries in the graph is less than or equal to a small value $s \ll d^2$. In particular, we use the Hamming distance $||\cdot||_H$ in calculating the distance of any two graphs.

**Definition 3.2** (Sparse Median Graph). *Let $\{\mathcal{G}^1, \ldots, \mathcal{G}^T\}$ be $T$ different binary graphs. The sparse median graph $\mathcal{G}_s^*$ is defined as*

$$\mathcal{G}_s^* := \underset{\mathcal{G} \in \{0,1\}^{d \times d}, ||\mathcal{G}||_H = s}{\operatorname{argmin}} \sum_{t=1}^{T} ||\mathcal{G} - \mathcal{G}^t||_H, \tag{3.2}$$

*where $||\cdot||_H$ represents the number of non-zero entries in the upper triangle of the matrix of interest.*

The next proposition presents an equivalent representation of $G_s^*$ and further discusses identifiablility conditions of the model.

**Proposition 3.3.** *Let $\mathcal{G}^t, t = 1, \ldots, T$ and $\mathcal{G}_s^*$ be the sparse median graph defined as above. Let $\zeta_{jk} = \sum_t \mathcal{G}_{jk}^t$ and $r_{jk}$ be the rank of all values $\{\zeta_{jk}\}_{j<k}$. Then:*

$$[\mathcal{G}_s^*]_{jk} = [\mathcal{G}_s^*]_{kj} = \begin{cases} 1, & \text{if } r_{jk} \leq s, \\ 0, & \text{if } r_{jk} > s. \end{cases}$$

*Moreover, the model is identifiable with respect to $\mathcal{G}_s^*$ if and only if there are no ties around rank $s$ for the sequence $\{\zeta_{jk}\}_{j<k}$.*

# 4  Methods

For $t = 1, \ldots, T$, let $\boldsymbol{x}_i^t = (x_{i1}^t, \ldots, x_{id}^t)^T, i = 1, \ldots, n_t$ be $n_t$ independent realizations of $\boldsymbol{X}_t$ (defined in Section 3.1). The observed data are $\{\boldsymbol{x}_i^t\}$ for $t = 1, \ldots, T$ and $i = 1, \ldots, n_t$ and the target is to estimate the sparse median graph $\mathcal{G}_s^*$ defined in Equation (3.2). The proposed method is a two step procedure. In the first step, the nonparanormal SKEPTIC is used to obtain the estimators $\{\widehat{\mathcal{G}}^t\}_{t=1}^T$ of $\{\mathcal{G}^t\}_{t=1}^T$. In the second step, $\mathcal{G}_s^*$ is estimated based on the estimators $\{\widehat{\mathcal{G}}^t\}_{t=1}^T$ obtained in the first step.

More specifically, in the first step, for each $t \in \{1, 2, \ldots, T\}$, let
$$\widehat{\mathbf{R}}_{jk} := \sin\left(\frac{\pi}{2}\widehat{\tau}_{jk}(\boldsymbol{x}_1^t, \ldots, \boldsymbol{x}_{n_t}^t)\right),$$
where $\widehat{\tau}_{jk}(\cdot)$ is defined in Equation (2.2). By using $\widehat{\mathbf{R}}^t = [\widehat{\mathbf{R}}_{jk}^t] \in \mathbb{R}^{d \times d}$ to estimate $\boldsymbol{\Sigma}^t$, one can plug $\widehat{\mathbf{R}}^t$ into CLIME to get estimates of $\boldsymbol{\Omega}^t$ and $\mathcal{G}^t$:
$$\widehat{\boldsymbol{\Omega}}^t = \arg\min_{\mathbf{M}} \sum_{j,k} |\mathbf{M}_{jk}|$$
$$\text{such that} \quad \|\widehat{\mathbf{R}}^t \mathbf{M} - \mathbf{I}_d\|_{\max} \leq \lambda_t, \tag{4.1}$$

where $\lambda_t > 0$ is a tuning parameter. Cai et al. (2011) show that this optimization can be decomposed into $d$ vector minimization problems, each of which can be reformulated as a linear program. Thus, it has the potential to scale to very large problems. Once $\widehat{\boldsymbol{\Omega}}^t$ is obtained, one can apply an additional thresholding step to estimate the graph, $\mathcal{G}^t$. For this, the graph estimator $\widehat{\mathcal{G}}^t \in \{0,1\}^{d \times d}$ is defined, in which a pair $(j,k)$ satisfies that $\widehat{\mathcal{G}}_{jk}^t \neq 0$ if and only if $\widehat{\boldsymbol{\Omega}}_{jk}^t > \gamma_t$. Here $\gamma_t$ is another tuning parameter. However, in practice, the CLIME algorithm works well without a second step truncation.

In the second step, provided that the estimates $\{\widehat{\mathcal{G}}^t, t = 1, \ldots, T\}$ have been obtained, the following equation is optimized to obtain $\widehat{\mathcal{G}}_s^*$
$$\widehat{\mathcal{G}}_s^* = \underset{\mathcal{G} \in \{0,1\}^{d \times d}, \|\mathcal{G}\|_H \leq s}{\arg\min} \sum_t \|\mathcal{G} - \widehat{\mathcal{G}}^t\|_H, \tag{4.2}$$

where the term $\|\mathcal{G}\|_H \leq s$ controls the sparsity degree of $\mathcal{G}$. In this paper, it is assumed that $s$ is known. Consider then the following proposition, which states that Equation (4.2) has a closed-form solution.

**Proposition 4.1.** *Let $\widehat{\zeta}_{jk}$ be defined as $\widehat{\zeta}_{jk} := \sum_t \widehat{\mathcal{G}}_{jk}^t$. Let $(j_1, k_1), (j_2, k_2), \ldots$ be $s$ pairs with the highest values in $\{\widehat{\zeta}_{jk}\}_{j<k}$. Then $\widehat{\mathcal{G}}_{jk} = 1$ if and only if $(j,k) \in \{(j_1, k_1), (j_2, k_2), \ldots\}$.*

**Remark 4.2.** *For simplicity, it is assumed that there are no ties around the rank $s$ for the sequence $\{\widehat{\zeta}_{jk}\}$. If the model discussed in Section 3 is identifiable and several mild conditions as shown in Section 5 hold, then there are no ties with high probability.*

# 5 Theoretical Properties

In this section the estimators from Section 4 are proved to be consistent for the true median graph. Notably, an nonasymptotic bound on the rate of convergence in estimating the sparse median graph with respect to the Hamming distance is provided.

Additional notation is required. Let $M_d$ be a quantity which may scale with the dimension $d$. Define

$$\mathcal{S}_d(q, s, M_d) := \left\{ \mathbf{\Omega} : ||\mathbf{\Omega}||_1 \leq M_d \text{ and } \max_{1 \leq j \leq d} \sum_{k=1}^{d} |\mathbf{\Omega}_{jk}|^q \leq s \right\}.$$

For $q = 0$, the class $\mathcal{S}_d(q, s, M_d)$ contains all the $s$-sparse matrices. The next theorem provides the parameter estimation and graph estimation consistency results for the nonparanormal SKEPTIC estimator defined in Equation (4.1).

**Theorem 5.1** (Liu et al. (2012)). *Let $\mathbf{X}^t \sim NPN_d(\mathbf{\Sigma}^t, f^t)$ with $\mathbf{\Omega}^t := [\mathbf{\Sigma}^t]^{-1} \in \mathbb{S}_d(q, s_t, M_d)$ with $0 \leq q < 1$. Let $\widehat{\mathbf{\Omega}}^t$ be defined in Equation (4.1). There exist constants, $C_0$ and $C_1$, only depending on $q$, such that whenever one chooses the tuning parameter $\lambda_t = C_0 M_d \sqrt{\frac{\log d}{n_t}}$, with probability no less than $1 - d^2$,*

$$||\widehat{\mathbf{\Omega}}^t - \mathbf{\Omega}^t||_2 \leq C_1 M_d^{2-2q} \cdot s \cdot \left( \frac{\log d}{n_t} \right)^{(1-q)/2}.$$

*Let $\widehat{\mathcal{G}}^t$ be the graph estimator defined in Section 4 with the second tuning parameter $\gamma_t := 4 M_d \lambda_t$. If it is further assumed that $\mathbf{\Omega} \in \mathcal{S}_d(0, s, M_d)$ and $\min_{j,k:\mathbf{\Omega}_{jk} \neq 0} |\mathbf{\Omega}_{jk}| \geq 2\gamma_t$, then*

$$\mathbb{P}(\widehat{\mathcal{G}}^t \neq \mathcal{G}^t) \leq 4 d^{-\epsilon_1},$$

*where $\epsilon_1 > 0$ is a constant that does not depend on $(n_t, d, s_t)$.*

*Proof.* Combining the Theorems 1 and 7 in Cai et al. (2011) and Theorem 4.2 in Liu et al. (2012). □

**Theorem 5.2** (Consistency). *With the above notation, the assumptions from Theorem 5.1, $\lambda_t, \gamma_t$ fixed and the model in Section 3 is identifiable, then*

$$\mathbb{P}(\widehat{\mathcal{G}}_s^* \neq \mathcal{G}_s^*) \leq 4 T d^{-\epsilon_1}, \tag{5.1}$$

*where $\widehat{\mathcal{G}}_s^*$ is defined as in Equation (4.2).*

*Proof.* If the model is identifiable, then one only needs to show that with high probability, all $\mathcal{G}^t$ can be recovered. Note that the union bound in Theorem 5.1 yields that

$$\mathbb{P}\left(\bigcup_{t=1}^{T}\{\widehat{\mathcal{G}}^t \neq \mathcal{G}^t\}\right) \leq \sum_{t=1}^{T}\mathbb{P}(\widehat{\mathcal{G}}^t \neq \mathcal{G}^t) \leq 4d^{-\epsilon_1} \leq 4Td^{-\epsilon_1}.$$

This completes the proof. □

The next theorem provides an upper bound of the rate of convergence with respect to the Hamming distance. Such a result is based on the recent explorations in graph recovery with respect to the Hamming distance (Ke et al., 2012; Jin et al., 2012).

**Theorem 5.3** (Rate of convergence). *Assume that the above assumptions in Theorems 5.1 and 5.2 hold. Let $\mathcal{A}_t$ be the event that*

$$\mathcal{A}_t := \{||\widehat{\mathcal{G}}^t - \mathcal{G}^t||_H \leq \delta_t\}$$

*and $\delta_t$ be defined such that $\mathbb{P}(\mathcal{A}_t) = 1 - o(d^{-\epsilon_2})$. Moreover, reorder $\{\zeta_{jk}\}_{j<k}$ to be $\zeta^{(1)} \geq \zeta^2 \geq \cdots \geq \zeta^{d(d-1)/2}$ and let $u^* = (\zeta^{(s)} - \zeta^{(s+1)})/2$. Then,*

$$\mathbb{P}\left(||\mathcal{G}_s^* - \widehat{\mathcal{G}}_s^*||_H \leq \frac{2\sum_{t=1}^{T}\delta_t}{u^*}\right) = 1 - o(Td^{-\epsilon_2}). \tag{5.2}$$

*Proof.* Let $\kappa^* := (\zeta^{(s)} + \zeta^{(s+1)})/2$. We reorder $\{\widehat{\zeta}_{jk}\}_{j<k}$ to be $\widehat{\zeta}^{(1)} \geq \widehat{\zeta}^{(2)} \geq \cdots \geq \widehat{\zeta}^{d(d-1)/2}$ and let $\widehat{\kappa}^* := (\widehat{\zeta}^{(s)} + \widehat{\zeta}^{(s+1)})/2$. Then $[\mathcal{G}_s^*]_{jk} \neq [\widehat{\mathcal{G}}_s^*]_{jk}$ if and only if

$$\text{sign}\left(\widehat{\zeta}_{jk} - \widehat{\kappa}^*\right) \cdot \text{sign}\left(\zeta_{jk} - \kappa^*\right) < 0.$$

Recall that $\delta_t \in \mathbb{R}$ is defined such that:

$$\mathbb{P}(\mathcal{A}_t) = \mathbb{P}(||\widehat{\mathcal{G}}^t - \mathcal{G}^t||_H > \delta_t) = o(d^{-\epsilon_2}). \tag{5.3}$$

Note that such a bound has been established in some constrained situations, for example, in Jin et al. (2012) (Theorem 1.2).

Let $\bar{\mathcal{G}}^*$ be the graph defined as:

$$\bar{\mathcal{G}}_{jk}^* = \bar{\mathcal{G}}_{kj}^* = \begin{cases} 1, & \text{if } \widehat{\zeta}_{jk} \geq \kappa^*, \\ 0, & \text{if } \widehat{\zeta}_{jk} < \kappa^*. \end{cases}$$

First we consider quantifying the difference between $\mathcal{G}_s^*$ and $\bar{\mathcal{G}}^*$. Let $u_{jk} := |\zeta_{jk} - \kappa^*|$. We reorder $\{u_{jk}\}$ from the smallest to the largest such that $u^{(1)} \leq u^{(2)} \leq \ldots \leq u^{(d(d-1)/2)}$. Let $N^*$ be defined as:

$$\sum_{t=1}^{N^*} u^{(t)} \leq \sum_{t}\delta_t \quad \text{and} \quad \sum_{t=1}^{N^*+1} u^{(t)} > \sum_{t}\delta_t.$$

Then, conditioning on the event $\bigcap_t \mathcal{A}_t$, we have the difference between $\widehat{\mathcal{G}}^t$ and $\mathcal{G}^t$ with regard to the Hamming distance is at most $\sum_t \delta_t$, and therefore

$$\|\mathcal{G}_s^* - \bar{\mathcal{G}}^*\|_H \leq N^*.$$

In particular, reminding that $u^* := (\zeta^{(s)} - \zeta^{(s+1)})/2 = \min_{(j,k)} u_{jk}$, we have

$$\|\mathcal{G}_s^* - \bar{\mathcal{G}}^*\|_H \leq \frac{\sum_{t=1}^T \delta_t}{u^*}. \tag{5.4}$$

Consider now quantifying the difference between $\bar{\mathcal{G}}^*$ and $\widehat{\mathcal{G}}_s^*$. Using Equation (5.4) and the fact that $\|\mathcal{G}_s^*\|_H = s$, one obtains

$$\min\left(0, s - \frac{\sum \delta_t}{u^*}\right) \leq \|\bar{\mathcal{G}}^*\|_H \leq s + \frac{\sum \delta_t}{u^*}.$$

Combing with the fact that $\|\widehat{\mathcal{G}}_s^*\|_H = s$, then

$$\|\widehat{\mathcal{G}}_s^* - \bar{\mathcal{G}}^*\|_H \leq \frac{\sum_{t=1}^T \delta_t}{u^*}.$$

Accordingly, by the triangle inequality, with high probability,

$$\|\widehat{\mathcal{G}}_s^* - \mathcal{G}_s^*\|_H \leq \|\widehat{\mathcal{G}}_s^* - \bar{\mathcal{G}}^*\|_H + \|\bar{\mathcal{G}}^* - \mathcal{G}_s^*\|_H \leq \frac{2\sum_{t=1}^T \delta_t}{u^*}.$$

This completes the proof.

□

**Remark 5.4.** *The bound constructed in Equation (5.2) is to balance the difference of $\{\mathcal{G}^t\}_{t=1}^T$ to $\mathcal{G}_s^*$ and the estimation error of $\widehat{\mathcal{G}}^t$ to $\mathcal{G}^t$. In other words, the better it is to differentiate $\{\mathcal{G}^t\}$ with $\mathcal{G}_s^*$ in the population level and the more accuracy $\widehat{\mathcal{G}}^t$ can approach $\mathcal{G}^t$, the better the final estimator can converge to the sparse median graph.*

# 6 Applications

In this section the performance of the proposed method is investigated on synthetic and real-world datasets. Three procedures are considered here:

NP: the algorithm to approximate the sparse median graphs using the "naive" Pearson sample correlation matrix on the whole datasets without considering the difference among different datasets.

Pearson: the algorithm to approximate the sparse median graphs using the Pearson sample correlation matrix on each datasets and then combining them together.

Kendall: the proposed rank-based estimator to approximate the sparse median graphs using the Kendall's tau correlation matrix on each datasets and then combining them together. Here the tuning parameters are selected by using the StARS stability-based approach (Liu

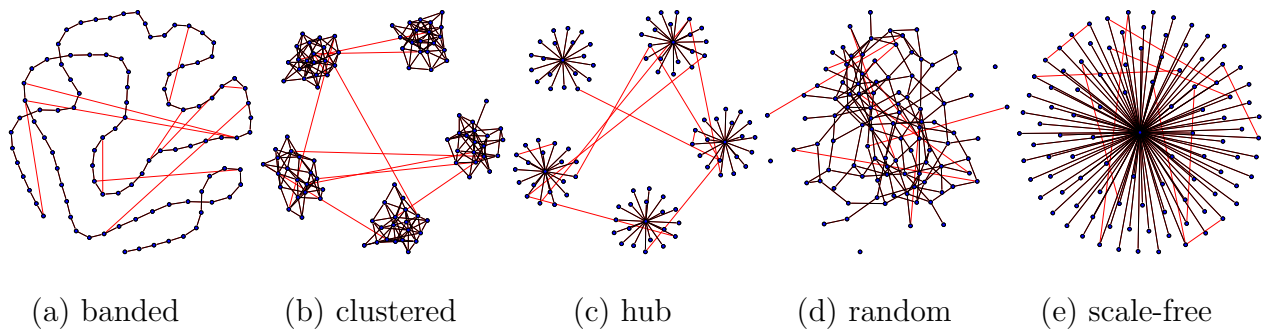

| (a) banded | (b) clustered | (c) hub | (d) random | (e) scale-free |

Figure 1: An illustration of the five graph patterns of the sparse median graphs $\mathcal{G}_s^*$ and the corresponding one individual dataset's graph $\mathcal{G}^t$. Here the black edges represent the ones present also in $\mathcal{G}_s^*$, and the red edges represent the one only present in $\mathcal{G}^t$.

et al., 2010).

## 6.1 Synthetic Data

In numeric simulations, there are $T = 10$ different datasets, $n = 100$ samples in each dataset, and the dimensionality is set to be $d = 100$. Each dataset is distributed to a different nonparanormal distribution, corresponding according to a different undirected graph $\mathcal{G}^t$. To build the model as developed in Section 3, assume that there exists a sparse median graph in each setting, denoted by $\mathcal{G}_s^*$. Then for $t = 1, \ldots, T$, 10 edges are randomly added to $\mathcal{G}_s^*$ for generating $\mathcal{G}^t$. The following five models are adopted for the graph, $\mathcal{G}_s^*$: banded, clustered, hub, random and scale-free. A typical run of the generated graphs $\mathcal{G}^t$ for a specific $t$ are illustrated in Figure 1. Here in each figure, the black edges represent the ones present in both $\mathcal{G}_s^*$ and $\mathcal{G}^t$, and the red edges represent the one only present in $\mathcal{G}^t$.

The covariance matrix, $\Sigma$, was then generated from the above five models using the huge package (Zhao et al., 2012). To generate $\Sigma^t$ for each submodel corresponding to each separate dataset, let $\Sigma^t_{jk} = \Sigma_{jk}$ for those $(j, k)$ such that $[\mathcal{G}_s^*]_{jk} = [\mathcal{G}^t]_{jk}$, and let $\Sigma^t_{jk} = 0.1$ for those not. Then for $t = 1, \ldots, T$, $n$ data points were generated $\boldsymbol{x}_1^t, \ldots, \boldsymbol{x}_n^t$ from $NPN_d(\Sigma^t, f)$. Here $f = \{f_1, \ldots, f_d\} = \{h_1, h_2, h_3, h_4, h_5, h_1, h_2, h_3, h_4, h_5, \ldots\}$, where

$$h_1^{-1}(x) := x, \ h_2^{-1}(x) := \frac{\text{sign}(x)|x|^{1/2}}{\sqrt{\int |t|\phi(t)dt}}, h_3^{-1}(x) := \frac{\Phi(x) - \int \Phi(t)\phi(t)dt}{\sqrt{\int \left(\Phi(y) - \int \Phi(t)\phi(t)dt\right)^2 \phi(y)dy}},$$

$$h_4^{-1}(x) := \frac{x^3}{\sqrt{\int t^6 \phi(t)dt}}, \ h_5^{-1}(x) := \frac{\exp(x) - \int \exp(t)\phi(t)dt}{\sqrt{\int \left(\exp(y) - \int \exp(t)\phi(t)dt\right)^2 \phi(y)dy}}.$$

Thus the data obtained is $\mathbf{X} = \{\boldsymbol{x}_1^1, \ldots, \boldsymbol{x}_n^1, \ldots, \boldsymbol{x}_1^T, \ldots, \boldsymbol{x}_n^T\}$. All three methods NP, Pearson and Kendall were used to approximate the sparse median graph $\mathcal{G}$. Results are presented in Figure 2.

There are two main conclusions drawn from the simulation results: (i) provided that one knows that the data are aggregated from different datasets, each of which is distributed differently, it is better to use the proposed algorithm than naively conducting analysis on the whole data without considering potential heteroscedasticity. This can be observed by comparing Pearson with NP. (ii) Kendall's empirical performance is significantly better than that of Pearson, which is not surprising because in the simulation set up every dataset is nonparanormally instead of Gaussian distributed. Nonetheless, the extent of the impact of errantly assuming Gaussianity can be severe.

## 6.2 ADHD Data

In this section the performance of the proposed method is investigated on a brain imaging dataset, the ADHD 200 dataset (Milham et al., 2012; Eloyan et al., 2012). The ADHD 200 dataset is a landmark study compiling over 1,000 functional and structural scans including subjects with and without attention deficit hyperactive disorder (ADHD). The data used in the analysis are from 776 subjects: 491 controls and 285 children diagnosed with ADHD of various subtypes. Each has at least one blood oxygen level dependent (BOLD) resting state functional MRI scans. The number of scans within an fMRI resting state session varies from 76 to 276, which were measured with different times between images (so called TR) as well as different scan lengths. The varying TR and length of scanning stress the importance of addressing subject-level heteroscedasticity in graph estimates. The data also include demographic variables as predictors. These include age, IQ, gender and handedness. We refer to Eloyan et al. (2012) for detailed data preprocessing procedures used in this analysis.

We constructed our predictors by extracting 264 voxels that broadly cover major functional regions of the cerebral cortex and cerebellum following Power et al. (2011). The locations of these 264 voxels are illustrated in Figure 3 and the value of each voxel is calculated as the mean of all data points inside these small seed regions. The information of the demographic variables was combined, in a final data matrix with the dimension $268 \times 776$. While different subjects will have different functional connectivity graphs, interest lies in common edges across different subjects' networks via classic neuroscientific principles of common functional specification and neural organization. The target is thus to use the concept of sparse median graphs and obtain sparse population averaged graphs. This was done separately for subjects with and without diagnosed ADHD. Several population median graph contrasts of interest were investigated and in include: ADHD case status (denoted by **Case** and **Control**), gender (denoted by **Female** and **Male**) and age. Given the pediatric population in the ADHD study, this investigates young adults versus children using a cutoff of 12 years. Subjects having ages larger than 12 years are denoted by **Senior** and those less than or equal to 12 years denoted by **Junior**.

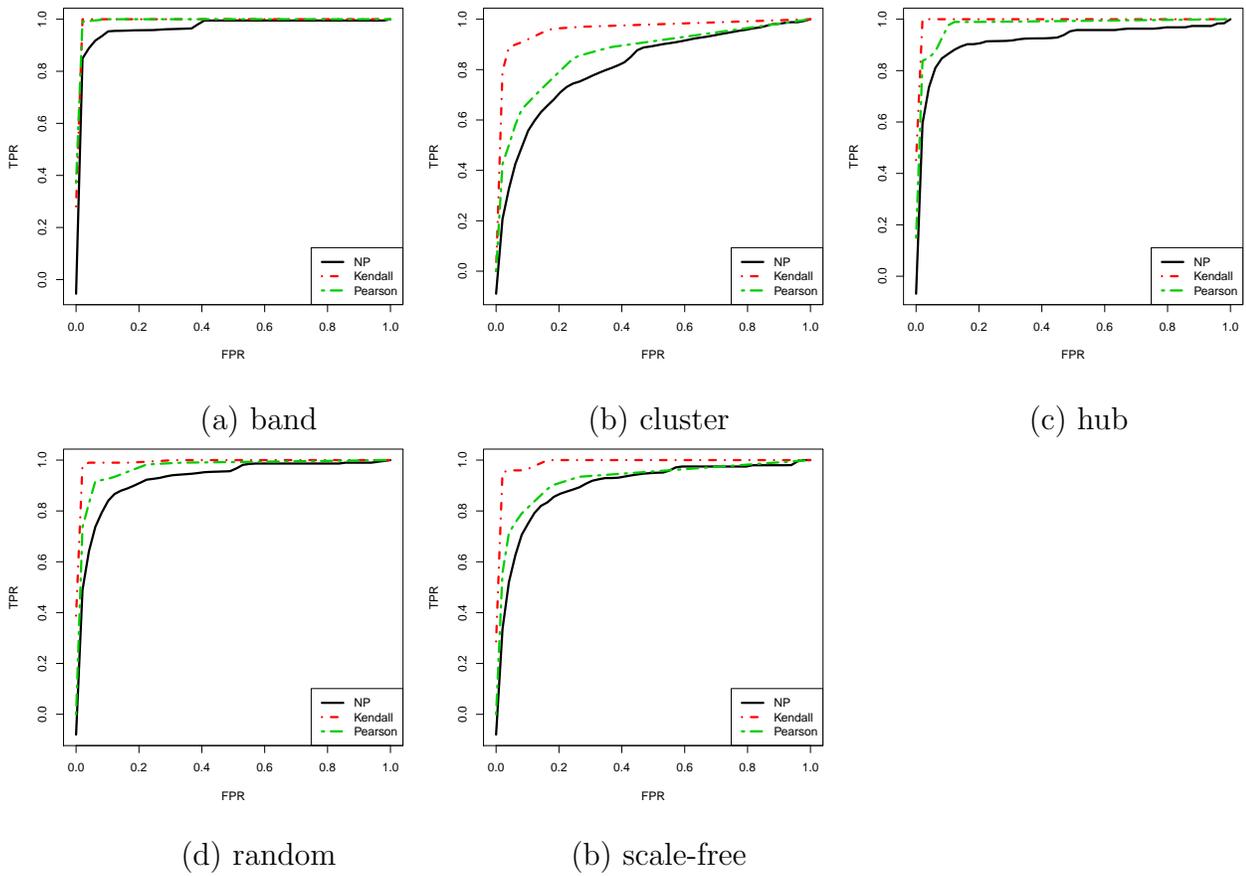

Figure 2: ROC curves in estimating the graphical models for different methods in five different graph patterns. Here $n = 100$ and $d = 100$.

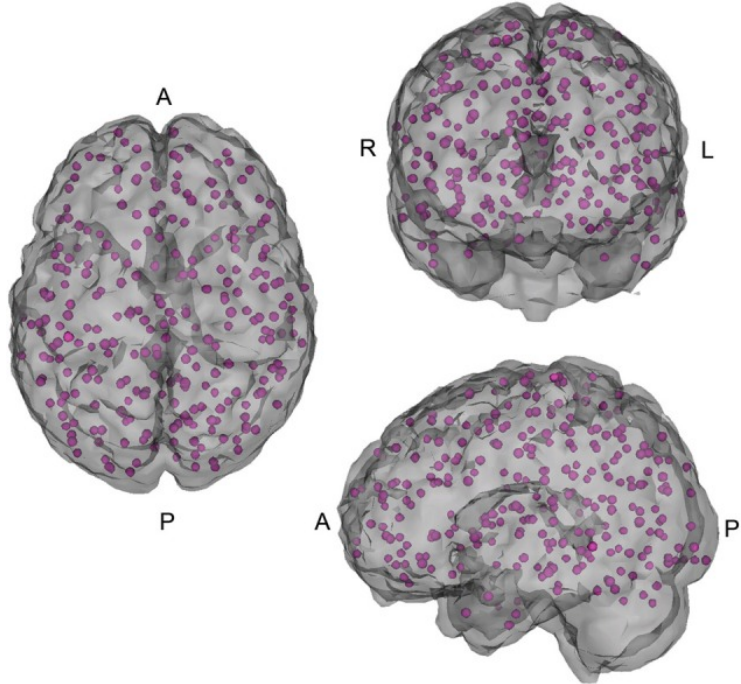

Figure 3: The illustration of the locations of the 264 nodes.

Consider the performances of Kendall and Pearson over the case and control data, regardless of gender and age. The graphs are shown in Figure 4. It can be observed that Kendall and Pearson provide very different sparse median graphs. Figure 5 further provides the comparison of the brain connectivity graphs obtained using Kendall on Case and Control data. Of note, there appears to be more homotopic connections (those connecting the left and right hemispheres) in the graph estimated using Kendall.

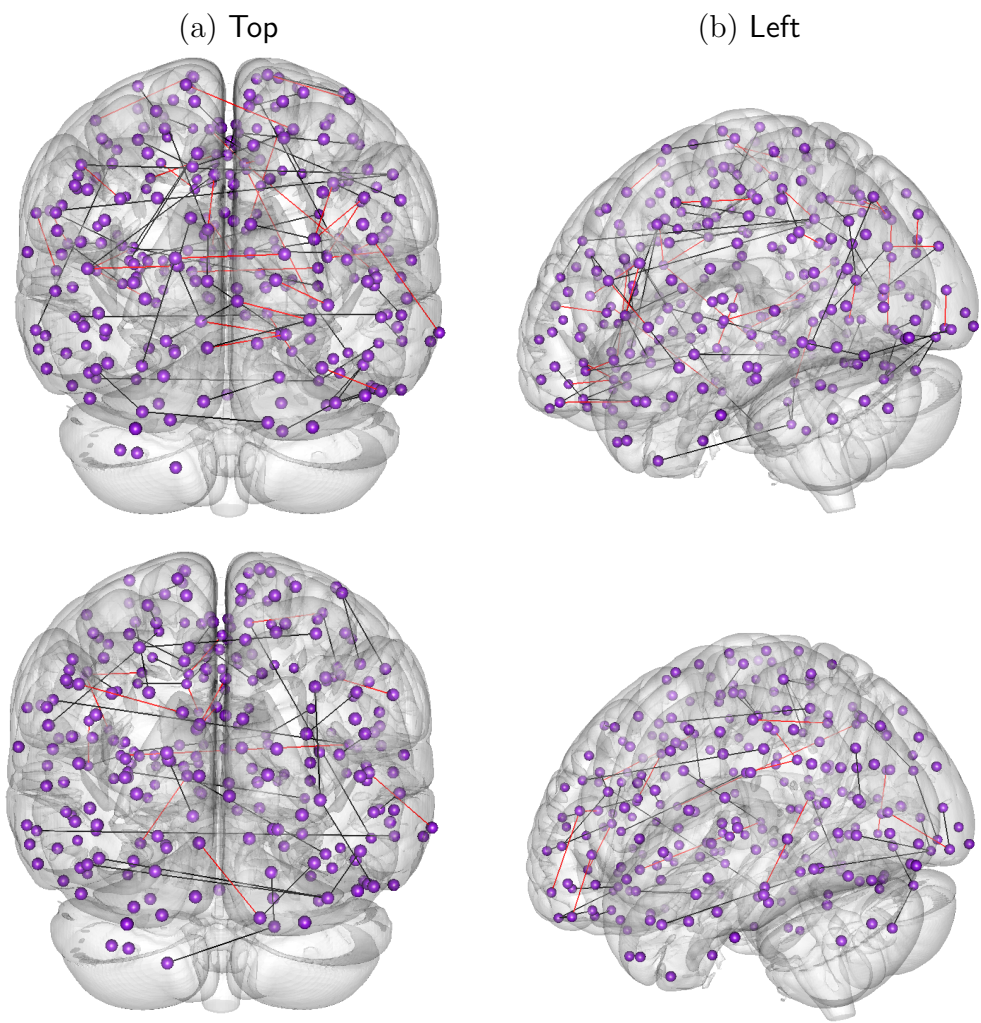

Figure 4: The difference of the estimated sparse median graphs using Kendall and Pearson for case and control subjects(top to bottom). Here the black color represents the edges only present in Kendall but not in Pearson, the red color represents the edges only present in Pearson but not in Kendall.

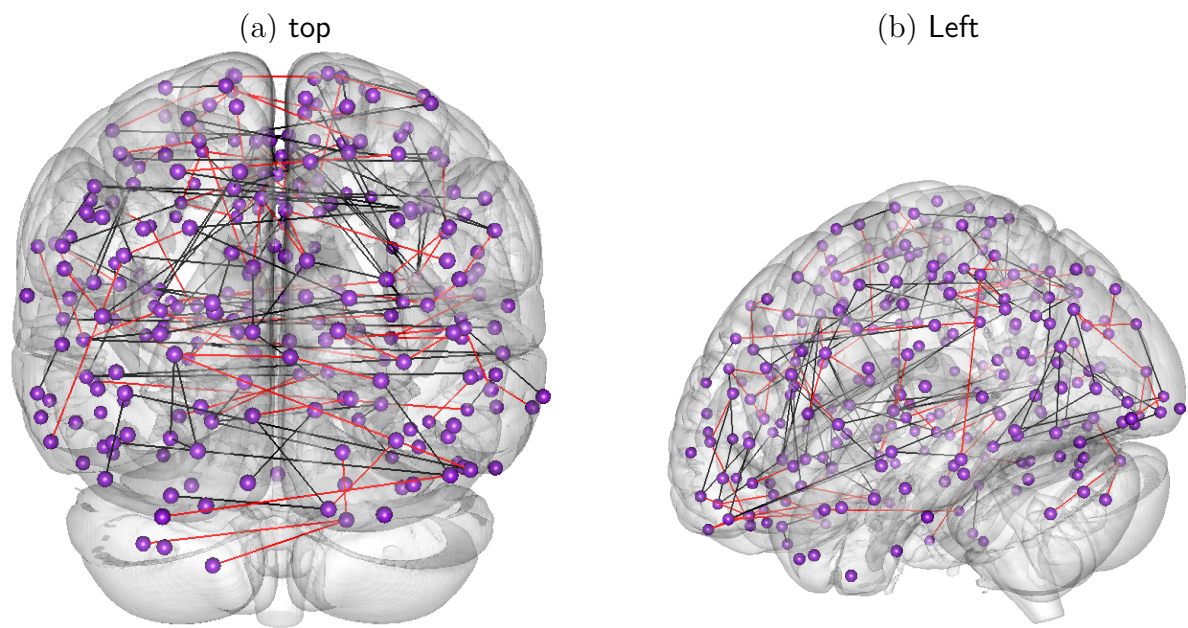

Figure 5: The difference between the estimated sparse median graphs of the cases and control subjects using Kendall. Here, the black color represents the edges only present in the median graph for cases but not in controls persons, while the red represents the opposite.

A more detailed analysis was subsequently attempted. The methods **Kendall** and **Pearson** are conducted on sub populations to find difference between graphs with different covariate levels. For example, median graphs between cases and controls were investigated within gender. Summary statistics for these subpopulations differences are presented in Table 1. It is observed that **pearson** tends to downgrade the difference, especially with regard to stratifying by age grouping.

Table 1: Summary statistics of the ADHD median graph differences estimated at whole data, gender and age levels. For any label $L$, $L$ represents the number of edges estimated using the data under this label. For any two labels $L_1$ and $L_2$, $L_1 > L_2$ represents the number of edges only present in the estimated graph of label $L_1$, but not in that of label $L_2$; $L_1 < L_2$ is vice versa. The label whole represents the entire data across all covariate levels.

|  | Kendall | | | | Pearson | | | |
|---|---|---|---|---|---|---|---|---|
| data | $A$ | $C$ | $A > C$ | $A < C$ | $A$ | $C$ | $A > C$ | $A < C$ |
| **Whole** | 946 | 962 | 79 | 95 | 971 | 988 | 85 | 102 |
| **Male** | 1010 | 979 | 171 | 140 | 1011 | 990 | 166 | 145 |
| **Female** | 979 | 975 | 110 | 106 | 1002 | 1005 | 112 | 115 |
| **Junior** | 947 | 952 | 110 | 115 | 961 | 968 | 92 | 99 |
| **Senior** | 1049 | 1053 | 139 | 143 | 1060 | 1060 | 112 | 112 |
| data | $M$ | $F$ | $M > F$ | $M < F$ | $M$ | $F$ | $M > F$ | $M < F$ |
| **Case** | 1010 | 979 | 169 | 138 | 1011 | 1002 | 158 | 149 |
| **Control** | 979 | 975 | 103 | 99 | 990 | 1005 | 94 | 109 |
| data | $J$ | $S$ | $J > S$ | $J < S$ | $J$ | $S$ | $J > S$ | $J < S$ |
| **Case** | 947 | 1049 | 137 | 239 | 961 | 1060 | 123 | 222 |
| **Control** | 952 | 1053 | 112 | 213 | 968 | 1060 | 93 | 185 |

## 6.3 ABIDE data

In this section we study the results of the proposed method on another brain imaging dataset, the Autism Brain Imaging Data Exchange (ABIDE) (Di Martino et al., 2013). The ABIDE data have a similar data structure as the ADHD data, and include 1,043 subjects, in which 499 are diagnosed autism. A similar data pre-processing approach was used, as discussed in the last section (Kang, 2013). Here the data are compressed by extracting 116 regions that are of interest from the AAL atlas (Tzourio-Mazoyer et al., 2002). Region labels are shown in Table 2. Refer to Di Martino et al. (2013) for detailed discussions in the data pre-processing procedures.

Table 2: Region information of the ABIDE data.

| region | description | region | description |
|---|---|---|---|
| 1 | Precentral-L | 59 | Parietal-Sup-L |
| 2 | Precentral-R | 60 | Parietal-Sup-R |
| 3 | Frontal-Sup-L | 61 | Parietal-Inf-L |
| 4 | Frontal-Sup-R | 62 | Parietal-Inf-R |
| 5 | Frontal-Sup-Orb-L | 63 | SupraMarginal-L |
| 6 | Frontal-Sup-Orb-R | 64 | SupraMarginal-R |
| 7 | Frontal-Mid-L | 65 | Angular-L |
| 8 | Frontal-Mid-R | 66 | Angular-R |
| 9 | Frontal-Mid-Orb-L | 67 | Precuneus-L |
| 10 | Frontal-Mid-Orb-R | 68 | Precuneus-R |
| 11 | Frontal-Inf-Oper-L | 69 | Paracentral-Lobule-L |
| 12 | Frontal-Inf-Oper-R | 70 | Paracentral-Lobule-R |
| 13 | Frontal-Inf-Tri-L | 71 | Caudate-L |
| 14 | Frontal-Inf-Tri-R | 72 | Caudate-R |
| 15 | Frontal-Inf-Orb-L | 73 | Putamen-L |
| 16 | Frontal-Inf-Orb-R | 74 | Putamen-R |
| 17 | Rolandic-Oper-L | 75 | Pallidum-L |
| 18 | Rolandic-Oper-R | 76 | Pallidum-R |
| 19 | Supp-Motor-Area-L | 77 | Thalamus-L |
| 20 | Supp-Motor-Area-R | 78 | Thalamus-R |
| 21 | Olfactory-L | 79 | Heschl-L |
| 22 | Olfactory-R | 80 | Heschl-R |
| 23 | Frontal-Sup-Medial-L | 81 | Temporal-Sup-L |
| 24 | Frontal-Sup-Medial-R | 82 | Temporal-Sup-R |
| 25 | Frontal-Med-Orb-L | 83 | Temporal-Pole-Sup-L |
| 26 | Frontal-Med-Orb-R | 84 | Temporal-Pole-Sup-R |
| 27 | Rectus-L | 85 | Temporal-Mid-L |
| 28 | Rectus-R | 86 | Temporal-Mid-R |
| 29 | Insula-L | 87 | Temporal-Pole-Mid-L |
| 30 | Insula-R | 88 | Temporal-Pole-Mid-R |
| 31 | Cingulum-Ant-L | 89 | Temporal-Inf-L |
| 32 | Cingulum-Ant-R | 90 | Temporal-Inf-R |
| 33 | Cingulum-Mid-L | 91 | Cerebelum-Crus1-L |
| 34 | Cingulum-Mid-R | 92 | Cerebelum-Crus1-R |
| 35 | Cingulum-Post-L | 93 | Cerebelum-Crus2-L |
| 36 | Cingulum-Post-R | 94 | Cerebelum-Crus2-R |
| 37 | Hippocampus-L | 95 | Cerebelum-3-L |
| 38 | Hippocampus-R | 96 | Cerebelum-3-R |
| 39 | ParaHippocampal-L | 97 | Cerebelum-4-5-L |
| 40 | ParaHippocampal-R | 98 | Cerebelum-4-5-R |
| 41 | Amygdala-L | 99 | Cerebelum-6-L |
| 42 | Amygdala-R | 100 | Cerebelum-6-R |
| 43 | Calcarine-L | 101 | Cerebelum-7b-L |
| 44 | Calcarine-R | 102 | Cerebelum-7b-R |
| 45 | Cuneus-L | 103 | Cerebelum-8-L |
| 46 | Cuneus-R | 104 | Cerebelum-8-R |
| 47 | Lingual-L | 105 | Cerebelum-9-L |
| 48 | Lingual-R | 106 | Cerebelum-9-R |
| 49 | Occipital-Sup-L | 107 | Cerebelum-10-L |
| 50 | Occipital-Sup-R | 108 | Cerebelum-10-R |
| 51 | Occipital-Mid-L | 109 | Vermis-1-2 |
| 52 | Occipital-Mid-R | 110 | Vermis-3 |
| 53 | Occipital-Inf-L | 111 | Vermis-4-5 |
| 54 | Occipital-Inf-R | 112 | Vermis-6 |
| 55 | Fusiform-L | 113 | Vermis-7 |
| 56 | Fusiform-R | 114 | Vermis-8 |
| 57 | Postcentral-L | 115 | Vermis-9 |
| 58 | Postcentral-R | 116 | Vermis-10 |

We then applied the two competing methods on this dataset. First consider the difference between the results produced by Pearson and Kendall. Table 3 provides some summary statistics about the sparse median graphs. It can be shown that Kendall more clearly differentiates disease status via the estimated sparse median graphs than Pearson. Of course the theoretical results lend further support, though the conclusion depends on the existence of actual differences between groups.

Table 3: Summary statistics of the ABIDE data networks estimated using the case and control data. Here a label indicator L in the table refers to the number of edges using the data in this label. L1>L2 represents the number of edges only present in the graph estimated using the data in L1, and vise versa.

| method | Control | Case | Control>Case | Control<Case |
|---|---|---|---|---|
| Pearson | 1033 | 1022 | 199 | 188 |
| Kendall | 948 | 945 | 202 | 199 |

We further visualize the results produced by Kendall. Figure 6(A) plots the sparse median graph on the control data and Figure 6(B) plots the difference between the graphs estimated using the control and case data. Each red line represents an edge that presents only in the case graph, while each black line represents an edge that presents only in the control graph. We find that in the case graph the probability of two remote regions linked together is higher than that in the control graph.

# 7 Discussion

In this paper we discuss the concept of median graph in analyzing complex aggregated datasets, propose a unified framework for conducting inference on such datasets and provide theoretical analysis and empirical experiments in evaluating the performance of the proposed methods. Experiments on both synthetic and real datasets illustrate the empirical usefulness of our models and methods.

The results on the ADHD 200 and ABIDE data sets offer a compelling preview of the potential utility of the approach. The current analysis is primarily illustrative, but motivates a more thorough inferential investigation of the median graph properties and network modification with disease.

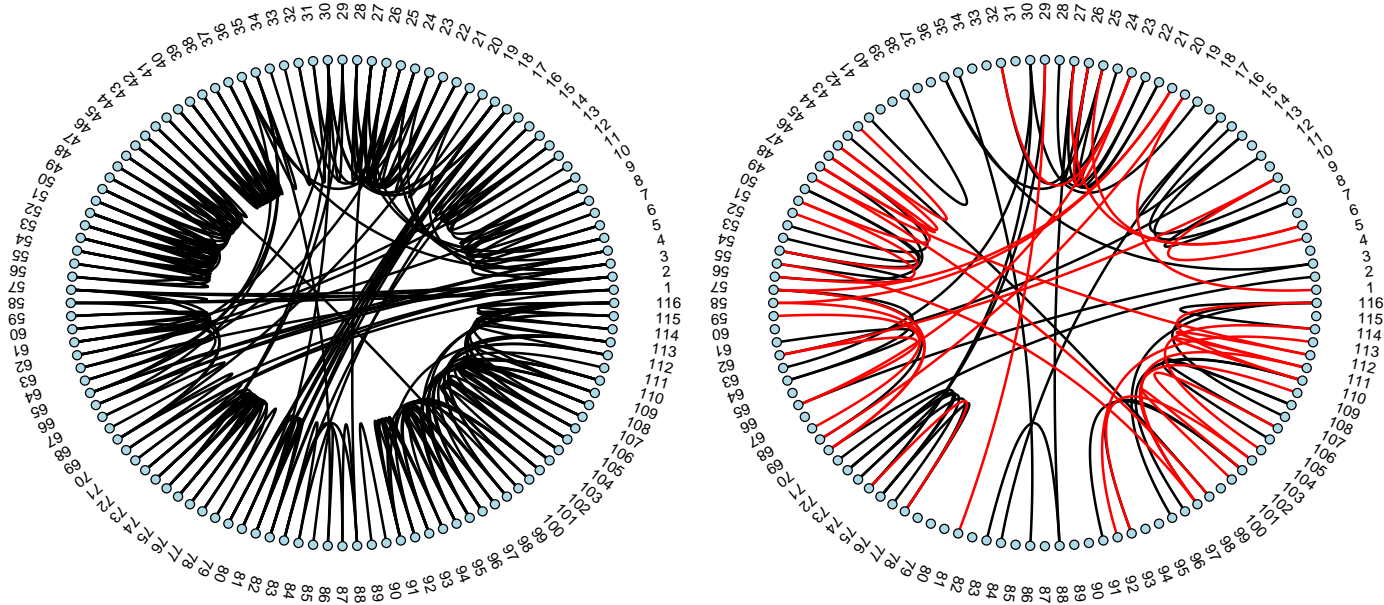

Figure 6: (A) The graph estimated by using the control data. (B) The difference between the estimated sparse median graphs of the case and control subjects using `kendall`. The black color represents the edges only present in controls but not in cases, while the red color represents the opposite. A truncation value 0.3 was used for visualization.

# References

Banerjee, O., Ghaoui, L. E., and d'Aspremont, A. (2008). Model selection through sparse maximum likelihood estimation. *Journal of Machine Learning Research*, 9:485–516.

Bühlmann, P. L. and van de Geer, S. A. (2011). *Statistics for High-Dimensional Data*. Springer.

Bullmore, E. and Sporns, O. (2009). Complex brain networks: graph theoretical analysis of structural and functional systems. *Nature Reviews Neuroscience*, 10(3):186–198.

Bunke, H. and Shearer, K. (1998). A graph distance metric based on the maximal common subgraph. *Pattern recognition letters*, 19(3):255–259.

Cai, T., Liu, W., and Luo, X. (2011). A constrained $\ell_1$ minimization approach to sparse precision matrix estimation. *Journal of the American Statistical Association*, 106(494):594–607.

Dempster, A. P. (1972). Covariance selection. *Biometrics*, 28(1):157–175.

Di Martino, A., Yan, C., Li, Q., Denio, E., Castellanos, F., Alaerts, K., Anderson, J., Assaf, M., Bookheimer, S., Dapretto, M., et al. (2013). The autism brain imaging data exchange: towards a large-scale evaluation of the intrinsic brain architecture in autism. *to appear Molecular psychiatry*.

Eloyan, A., Muschelli, J., Nebel, M., Liu, H., Han, F., Zhao, T., Barber, A., Joel, S., Pekar, J., Mostofsky, S., et al. (2012). Automated diagnoses of attention deficit hyperactive disorder using magnetic resonance imaging. *Frontiers in Systems Neuroscience*, 6:61.

Fingelkurts, A., Kähkönen, S., et al. (2005). Functional connectivity in the brain–is it an elusive concept? *Neuroscience and biobehavioral reviews*, 28(8):827–836.

Friedman, J. H., Hastie, T., and Tibshirani, R. (2007). Sparse inverse covariance estimation with the graphical lasso. *Biostatistics*, 9(3):432–441.

Friston, K. (2011). Functional and effective connectivity: a review. *Brain Connectivity*, 1(1):13–36.

Horwitz, B. et al. (2003). The elusive concept of brain connectivity. *Neuroimage*, 19(2):466–470.

Hsieh, C.-J., Sustik, M. A., Ravikumar, P., and Dhillon, I. S. (2011). Sparse inverse covariance matrix estimation using quadratic approximation. In *Advances in Neural Information Processing Systems (NIPS)*, volume 24.

Jiang, X., Munger, A., and Bunke, H. (2001). An median graphs: properties, algorithms, and applications. *Pattern Analysis and Machine Intelligence, IEEE Transactions on*, 23(10):1144–1151.

Jin, J., Zhang, C., and Zhang, Q. (2012). Optimality of graphlet screening in high dimensional variable selection. *arXiv preprint arXiv:1204.6452*.

Kang, J. (2013). Abide data preprocessing. *personal communication*.

Ke, T., Jin, J., and Fan, J. (2012). Covariance assisted screening and estimation. *arXiv preprint arXiv:1205.4645*.

Lam, C. and Fan, J. (2009). Sparsistency and rates of convergence in large covariance matrix estimation. *Annals of Statistics*, 37(6B):42–54.

Li, L. and Toh, K.-C. (2010). An inexact interior point method for $\ell_1$-reguarlized sparse covariance selection. *Mathematical Programming Computation*, 2(3):291–315.

Liu, H., Han, F., Yuan, M., Lafferty, J., and Wasserman, L. (2012). High dimensional semiparametric gaussian copula graphical models. *Annals of Statistics*, 40(4):2293–2326.

Liu, H., Lafferty, J., and Wasserman, L. (2009). The nonparanormal: Semiparametric estimation of high dimensional undirected graphs. *The Journal of Machine Learning Research*, 10:2295–2328.

Liu, H., Roeder, K., and Wasserman, L. (2010). Stability approach to regularization selection (stars) for high dimensional graphical models. In *Advances in Neural Information Processing Systems*, volume 23.

Liu, I.-M. and Agresti, A. (1996). Mantel-haenszel-type inference for cumulative odds ratios with a stratified ordinal response. *Biometrics*, 52(4):1223–1234.

Liu, W. and Luo, X. (2012). High-dimensional sparse precision matrix estimation via sparse column inverse operator. *arXiv preprint arXiv:1203.3896*.

Meinshausen, N. and Bühlmann, P. (2006). High-dimensional graphs and variable selection with the lasso. *The Annals of Statistics*, 34(3):1436–1462.

Milham, M. P., Fair, D., Mennes, M., and Mostofsky, S. H. (2012). The ADHD-200 consortium: a model to advance the translational potential of neuroimaging in clinical neuroscience. *Frontiers in Systems Neuroscience*, 6:62.

Peng, J., Wang, P., Zhou, N., and Zhu, J. (2009). Partial correlation estimation by joint sparse regression models. *Journal of the American Statistical Association*, 104(486):735–746.

Power, J., Cohen, A., Nelson, S., Wig, G., Barnes, K., Church, J., Vogel, A., Laumann, T., Miezin, F., Schlaggar, B., et al. (2011). Functional network organization of the human brain. *Neuron*, 72(4):665–678.

Ravikumar, P., Wainwright, M., Raskutti, G., and Yu, B. (2009). Model selection in Gaussian graphical models: High-dimensional consistency of $\ell_1$-regularized MLE. In *Advances in Neural Information Processing Systems*, volume 22.

Rothman, A. J., Bickel, P. J., Levina, E., and Zhu, J. (2008). Sparse permutation invariant covariance estimation. *Electronic Journal of Statistics*, 2:494–515.

Rubinov, M. and Sporns, O. (2010). Complex network measures of brain connectivity: uses and interpretations. *Neuroimage*, 52(3):1059–1069.

Scheinberg, K., Ma, S., and Glodfarb, D. (2010). Sparse inverse covariance selection via alternating linearization methods. In *Advances in Neural Information Processing Systems (NIPS)*, volume 23.

Tzourio-Mazoyer, N., Landeau, B., Papathanassiou, D., Crivello, F., Etard, O., Delcroix, N., Mazoyer, B., and Joliot, M. (2002). Automated anatomical labeling of activations in SPM using a macroscopic anatomical parcellation of the MNI MRI single-subject brain. *Neuroimage*, 15(1):273–289.

Xue, L. and Zou, H. (2012). Regularized rank-based estimation of high-dimensional non-paranormal graphical models. *The Annals of Statistics*, 40(5):2541–2571.

Yuan, M. (2010). High dimensional inverse covariance matrix estimation via linear programming. *The Journal of Machine Learning Research*, 11:2261–2286.

Zhao, T., Liu, H., Roeder, K., Lafferty, J., and Wasserman, L. (2012). The huge package for high-dimensional undirected graph estimation in r. *The Journal of Machine Learning Research*, 13:1059–1062.